# Differentiating Workload using Pilot's Stick Input in a Virtual Reality Flight Task

Evy van Weelden, Carl W. E. van Beek, Maryam Alimardani, Travis J. Wiltshire, Wietse D. Ledegang, Eric L. Groen, and Max M. Louwerse

*Abstract*—**High-risk operational tasks such as those in aviation require training environments that are realistic and capable of inducing high levels of workload. Virtual Reality (VR) offers a simulated 3D environment for immersive, safe and valid training of pilots. An added advantage of such training environments is that they can be personalized to enhance learning, e.g., by adapting the simulation to the user's workload in real-time. The question remains how to reliably and robustly measure a pilot's workload during the training. In this study, six novice military pilots (average of 34.33 flight hours) conducted a speed change maneuver in a VR flight simulator. In half of the runs an auditory 2-back task was added as a secondary task. This led to trials of low and high workload which we compared using the pilot's control input in longitudinal (i.e., pitch) and lateral (i.e., roll) directions. We extracted Pilot Inceptor Workload (PIW) from the stick data and conducted a binary logistic regression to determine whether PIW is predictive of task-induced workload. The results show that inputs on the stick along its longitudinal direction were predictive of workload (low vs. high) when performing a speed change maneuver. Given that PIW may be a task-specific measure, future work may consider (neuro)physiological predictors. Nonetheless, the current paper provides evidence that measuring PIW in a VR flight simulator yields real-time and non-invasive means to determine workload.**

## I. INTRODUCTION

IN high-risk operational environments such as aviation, an increased workload could adversely affect pilot ability and flight performance [1], [2], [3], and therefore overall flight safety. Aviation training programs often employ simulated tasks with increased workload to prepare pilots for demanding situations in real flight. However, extremely high or low levels of workload have shown to negatively affect one's training [4].

Training pilots using real aircraft is costly and time-consuming. To resolve this problem, Virtual Reality (VR) simulations have demonstrated to be a safe, realistic and cost-effective training environment to complement pilot training programs [5]–[7]. VR training systems have shown ecological validity, eliciting similar performance and patterns of brain and cardiac activity in a variety of application domains, including aviation [5], [8]–[10].

In simulated flights, increase in workload has been associated with increased deviations from targeted flight parameters [11], [12]. Behavioral measures that target workload are of importance in the understanding of learning [13], and adaptive human-controlled systems [14]. Hebbar and Pashilkar [15] argued that these parameter deviations do not sufficiently reflect changes in workload on their own, as pilots are generally able to perform well under demanding conditions. While performance measures are often associated with workload, people can experience a variety of workload levels while maintaining the same level of performance, for example, due to (in)experience and task awareness [16], [17], but also due to task characteristics such as task duration and the required level of information processing [18]. Therefore, to reliably assess a pilot's workload, Hebbar and Pashilkar [15] suggested to quantify pilot control over the inceptors in the (simulator) cockpit, such as the stick. Stick control serves as a direct measure of the human operation and interaction with the (virtual) aircraft and hence provides valuable information in addition to flight performance measures such as deviations from given parameters or flight paths.

The stick control is generally described using two measures: Duty Cycle and Aggressiveness [15], [19]. Duty Cycle refers to the amount of time the controller changes the input on the stick (in percentage), and Aggressiveness indicates the rate of change of the stick movements [12], [19]. Using these two measures allows for estimating Pilot Inceptor Workload (PIW): a two-dimensional measure of pilot gain, i.e., the effort put into aircraft control [19]–[22]. This concept has been schematically outlined in Fig. 1. PIW can also be transformed into a one-dimensional measure [19], where Duty Cycle and Aggressiveness are merged into one PIW value. This one-dimensional PIW allows for direct comparisons to other pilot gain and flight performance measures [19], [20].

Manuscript received ... The research reported in this study is funded by the MasterMinds project, part of the RegionDeal Mid- and West-Brabant, and is co-funded by the Ministry of Economic Affairs and Municipality of Tilburg awarded to MML. *(Corresponding author: E. van Weelden).*

Evy van Weelden, Carl W. E. van Beek, Maryam Alimardani, Travis J. Wiltshire, and Max M. Louwerse are with Tilburg University, Tilburg, NL, at the department of Cognitive Science and Artificial Intelligence (e-mail: e.vanweelden@tilburguniversity.edu; c.w.e.vanbeek@tilburguniversity.edu; m.alimardani@tilburguniversity.edu; t.j.wiltshire@tilburguniversity; m.m.louwerse@tilburguniversity.edu). Wietse D. Ledegang and Eric L. Groen are with TNO, Soesterberg, NL (e-mail: wietse.ledegang@tno.nl; eric.groen@tno.nl).

This work involved human subjects in its research. Approval of all ethical and experimental procedures and protocols was granted by the Research Ethics Committee of Tilburg School of Humanities and Digital Sciences (Application No. REDC2021.36). This manuscript has Supplementary Materials.

Color versions of one or more of the figures in this article are available online at http://ieeexplore.ieee.org.

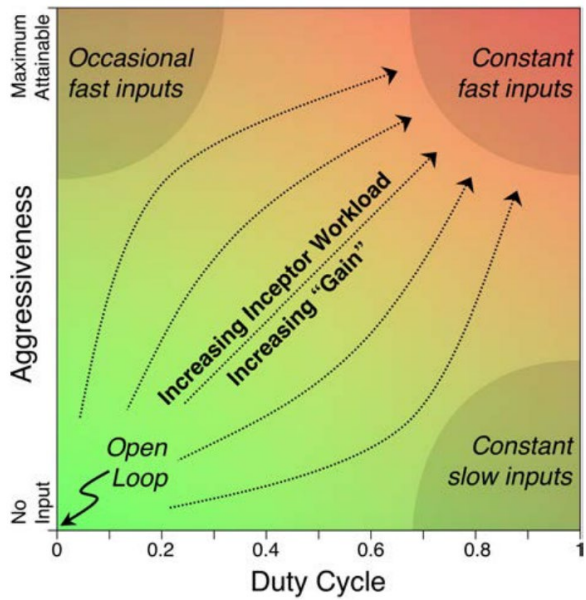

**Fig. 1.** Pilot Inceptor Workload. Reprinted from [22]. There is a hypothetical linear relationship between Duty Cycle and Aggressiveness, in which higher values are associated with increased workload [22].

Previously, Rajshekar Reddy et al. [23] attempted to estimate the workload of non-pilot participants using PIW and physiological measures in a VR flight simulator, when flying a combination of six maneuvers. They found that Duty Cycle and Aggressiveness differed between the tasks, and observed that Aggressiveness correlated with subjective ratings of workload, while Duty Cycle correlated with the number of eye saccades. Based on these findings, [23] concluded that a two-dimensional measure of PIW can give an indication of the complexity of the flight maneuvers and hence can be employed in evaluating pilots' workload and performance in a non-invasive manner. However, more research is required to further understand the impact of workload manipulations on two-dimensional (i.e., Duty Cycle and Aggressiveness) as well as one-dimensional forms of PIW, for example, by studying the relationship between mental workload and PIW within-tasks rather than using different maneuvering tasks for PIW measurement.

The current study investigated the efficacy of using stick data from a small homogeneous group of trained military pilots to quantify two levels of workload during VR-based flight tasks. We employed speed change as the primary flight task. Each pilot conducted 12 trials of the speed change task where workload in half of the trials was manipulated by adding a secondary task which was an auditory 2-back task [24]. Using this design, we intended to measure changes in PIW that could be fully attributed to the changes of workload, rather than task characteristics as in [23]. To do this, we extracted Duty Cycle, Aggressiveness, and one-dimensional PIW from the stick data, and compared these measures between low and high workload trials. Additionally, we used one-dimensional PIW to predict the level of workload. The ultimate aim of this study is to validate whether PIW has the potential to be used in real-time monitoring of pilot's workload for the purpose of adaptive flight training, for instance by providing adjusted task difficulty when a state of under- and/or overload is detected.

## II. METHODS

### A. Participants

Six male Royal Netherlands Air Force student pilots ($M_{age}$ = 25.00, $SD_{age}$ = 6.36) who just finished elementary training for fixed wing aircraft on a Pilatus PC-7 with an average of 34.33 flight hours ($SD$ = 4.50) participated in this study. Informed consent was collected from each pilot prior to the experiment. The sample size is relatively small, but we have compensated with a high number of trials per subject (cf., section II. C. Procedure). It is common within the field of aviation to have a small sample size, due to limited access to resources, such as skilled pilots, instructors, and simulator time (e.g., [9], [20], [21], [23]). Besides, our sample is very homogeneous, as the subject-level variation in the dataset is (near) zero (cf., section II. E. Statistical analysis).

This study was approved by the Tilburg School of Humanities and Digital Sciences Research Ethics and Data Management Committee of Tilburg University and the TNO institutional ethics committee, and was in accordance with the (revised) Helsinki Declaration.

### B. Materials

A VR flight simulator of the Pilatus PC-7 aircraft (multiSIM B.V., the Netherlands) was used, including an inceptor (stick) with control loading (i.e., providing forces on the stick related to the simulated flight dynamics and trim as it would in real flight), see Fig. 2A and B. The VR content was displayed to the pilots using the Varjo Aero head-mounted display (HMD).[1] Flight parameters and pilots' input on the flight controls, such as the stick, were recorded in flight logs with varying sample rates up to 500Hz.

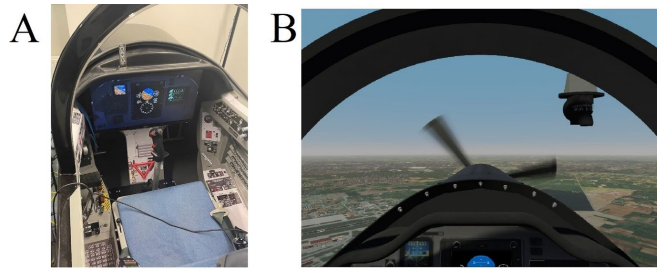

**Fig. 2.** Experimental setup. (A) Participants sat in a PC-7 cockpit mock-up featuring a stick and pedals with control loading, and a realistic throttle with yaw-trim controls, (B) participants viewed a simulated VR environment while performing the aviation task.

---

[1] A wireless 32-channel EEG system (g.Nautilus PRO, g.tec medical engineering GmbH, Austria) was used to record brain activity during the simulated flight tasks. However, given the focus on stick data in this paper, EEG results fall outside the scope of the current study.

## C. Procedure

A pre-experimental questionnaire collected participant descriptives (sex, age, handedness, VR and flight experience) and VR training expectancies based on the Unified Theory of Acceptance and Use of Technology (UTAUT) [25], [26]. The post-experimental questionnaire included the UTAUT as well as the System Usability Questionnaire (SUS) [27], and the ITC-Sense of Presence Inventory (ITC-SOPI) [28]. The UTAUT, SUS and ITC-SOPI questionnaires were used to inquire about the subjective user experience of the VR simulation.

Prior to this experiment, the participants were exposed to another study in which they performed three basic flight maneuvers (i.e., straight-and-level, level turn, and a speed change), using the same VR-simulator setup. Each maneuver was repeated three times in trials of 210 seconds each, followed by a fourth run in which the maneuver was performed while simultaneously executing an additive N-back task (2-back) as a measure of cognitive spare capacity. The applied N-back task required the participant to remember the last two letters of an auditory sequence of continuously changing letters at a fixed 3-seconds interval with a 25% repetition probability. The participant was instructed to make a self-paced response by pressing a dedicated button on the throttle if the letter heard was identical to the letter two trials back and to withhold a response if the letter was different. Regardless of the flight maneuver, participants also had to perform a lookout task to detect visual objects that semi-randomly appeared in the outside environment. The results of this study are described in [29].

After completing this preceding study and a short break, the participants were introduced to the present study, consisting of twelve trials of the same speed change task. During the speed change tasks, each starting from a trimmed condition at 5,000ft altitude, participants had to decelerate from 180 to 110 Knots Indicated Air Speed (KIAS), while maintaining altitude, constant heading and coordinated flight.

The test phase consisted of twelve trials of a speed change task (change speed from 180 knots 110 knots). Half of the twelve trials included the additional auditory N-back task (as described previously) to induce a higher level of workload [30]. Randomized sequences of trials were computer-generated with the use of the MATLAB R2022b sample() function [31], according to the block randomization method (see description in Supplementary Materials). Fig. 3 displays a schematic outline of the procedure. The type of trial ('N-back' or 'No N-back') was communicated to the participants prior to the onsets of each trial. Responses to the N-back task were acquired by button presses on the throttle.

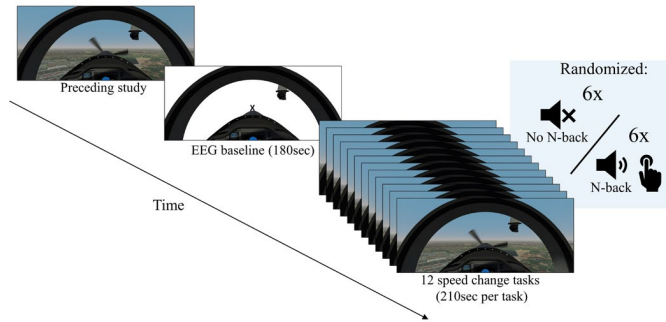

**Fig. 3.** Schematic task procedure. Six of the twelve trials had an additional auditory N-back task (2-back).

## D. Data processing and analysis

The processing of flight logs was performed in MATLAB R2022b using the Signal Processing Toolbox [31]. First, the stick data was resampled to 100 Hz for consistency, as the sample rate varied and could range up to 500 Hz during logging.

*1) Calculating Duty Cycle and Aggressiveness:* To obtain PIW, Duty Cycle and Aggressiveness were calculated for each trial across two axes; the longitudinal (i.e., forward-aft) and lateral axis (i.e., sideward), according to equation (1) and (2) respectively [19], [21].

$$Duty\ Cycle = \frac{1}{t_n - t_2} \sum_{i=2}^{n} x_i$$

$$x_i = \begin{cases} 0, & \left|\frac{\delta_i - \delta_{i-1}}{t_i - t_{i-1}}\right| < thr\ and\ |\delta_i| < \delta_{max} \\ 1 \end{cases} \quad (1)$$

$$Aggressiveness = \sqrt{\frac{1}{n-1} \sum_{i=2}^{n} \left(\frac{\delta_i - \delta_{i-2}}{t_i - t_{i-1}}\right)^2} \quad (2)$$

in which $t_i$ indicates the sample number of a trial with the length of $n$ samples, and $\delta$ is the amount of stick deflection with $\delta_{max}$ indicating the maximum deflection. Our noise threshold (*thr*) was set to zero in (1) as machine or aircraft vibrations were not part of our simulator set-up and it was assumed that all stick control inputs were voluntary.

*2) Calculating one-dimensional Pilot Inceptor Workload:* Although four different formulas are used in literature to compute the one-dimensional PIW (PIW1) [18], for easy interpretation of the outcomes we decided to use the following formula (3):

$$PIW1 = \sqrt{agg * dc} \quad (3)$$

in which *agg* indicates normalized Aggressiveness, and *dc* is Duty Cycle. Aggressiveness was normalized using the inverse of the exponential function between Duty Cycle and Aggressiveness [19]. Altogether, equations (1) to (3) resulted in three measures of stick control (i.e., Duty Cycle, Aggressiveness and PIW1) per axis (i.e., longitudinal and lateral stick data).

See Supplementary Materials and code repository (https://github.com/evyvanweelden/PilotInceptorWorkload.git) for implementation.

## E. Statistical analysis

Statistical analyses were performed in R [32]. Lateral and Longitudinal measures of Duty Cycle, Aggressiveness, and PIW1 were tested for normality using Shapiro-Wilk tests. Variables that were not normally distributed, were tested for differences between conditions using the non-parametric Wilcoxon signed-rank test. Normally distributed variables were analyzed using dependent t-tests.

Outliers were identified using the is_outlier() function of the rstatix package [33]. Four datapoints were marked as outliers, but none of these were extreme outliers (i.e., they were not greater than three times the interquartile range from the upper quartile). We decided not to remove any outliers due to the observation that their removal did not meaningfully change the results of our comparison tests, and the fact that there were no influential data points to our predictive model.

In order to assess the suitability of the data for a binary logistic regression model (i.e., a regression model in which the outcome variable is dichotomous), we examined whether there was an effect of trial number on the outcome, checked variables for high redundancy, and evaluated multicollinearity.

First, we looked whether there was an effect of trial number on PIW1 using the lm() function in R. There was no linear effect of trial number on lateral nor longitudinal PIW1.

Then, we examined a correlation matrix as a first attempt to examine multicollinearity and to check what variables correlated highly with each other (see Supplementary Materials). It appeared that all variables originating from the longitudinal data (i.e., Duty Cycle, Aggressiveness and PIW1) correlated with each other. The same pattern of correlations was also found for the lateral variables. We therefore decided to only include lateral and longitudinal PIW1 in the model as they reflected both Duty Cycle and Aggressiveness within their plane of motion. We calculated variance inflation factor (VIF) to confirm that there was no multicollinearity between lateral and longitudinal PIW1 (see Supplementary Materials for a description and the resulting VIF values). Due to a random effect variance of zero in the model, no random factor for participant was required [34].

We next conducted the binary logistic regression using the glm() function from R. This analysis was performed to assess whether PIW is predictive of workload, and what the relationship is between multiple predictors (i.e., measures of stick control) and the outcome variable (i.e., level of workload). The model included lateral PIW1 and longitudinal PIW1 as predictors of (Low and High) Workload.

III. RESULTS

To verify that the N-back task was attended to during the speed change task, accuracy scores per trial were computed. The overall accuracy of the N-back task averaged over all participants was .87 ($SD$ = .15), demonstrating all participants put effort into the additional task.

A. Subjective user experience

The outcomes of the UTAUT, SUS and ITC-SOPI questionnaires are shown in Table I. No statistical tests were conducted on the questionnaire data due to data sparsity. The anecdotal evidence of the outcomes show that behavioral intention (UTAUT) to use the VR flight training system increased after the experience, and system usability (SUS) was rated relatively high by our small sample of pilots ($M$ = 76.25 out of a maximum possible score of 100).

B. Comparisons of Duty Cycle and Aggressiveness

The results, presented in Table II, show that both the Duty Cycle and Aggressiveness in longitudinal direction (i.e., pitch control input) were significantly higher in the High Workload condition than in the Low Workload condition. This indicates that participants showed more variations in pitch direction when executing the speed change task with an additional N-back task. Accordingly, the difference between workload conditions is visible in the longitudinal PIW plot (Fig. 4A), and less so in the lateral PIW plot (Fig. 4B).

TABLE I
Questionnaire outcomes

| Question-naire | Sub-scale | Mean(SD) score obtained | |
| --- | --- | --- | --- |
| | | Pre-experimental | Post-experimental |
| UTAUT | Performance expectancy | 3.83 (.52) | 3.79 (.56) |
| | Effort expectancy | 4.04 (.60) | 4.42 (.49) |
| | Social influence | 3.46 (.40) | 3.38 (.90) |
| | Facilitating conditions | 3.50 (.35) | 3.46 (.37) |
| | Behavioral intention | 3.67 (.67) | 4.00 (.60) |
| SUS | | | 76.25 (6.28) |
| ITC-SOPI | Engagement | | 3.27 (.24) |
| | Presence | | 3.57 (.28) |
| | Negative effects | | 2.53 (.69) |
| | Naturalness | | 3.72 (.27)[1] |

*Note*. [1] ITC-SOPI Naturalness shows average score of five participants, because of a missing value for one participant.

C. Comparisons of one-dimensional Pilot Inceptor Workload

As shown in Table II, PIW1 in longitudinal direction (i.e., pitch control input) was significantly higher in the High Workload trials when compared to the Low Workload trials. At the same time, PIW1 did not differ between the Low and High Workload conditions in the lateral direction (i.e., roll control input). Similar to Duty Cycle and Aggressiveness, our one-dimensional measure of stick control input increased when the N-back task was added to trials, only for stick control in the longitudinal direction. See Fig. 5 for boxplots with the datapoints per participant.

TABLE II
Test Statistics for Normally Distributed Data and Non-Parametric Data

| Normally Distributed Data | | | | | |
|---|---|---|---|---|---|
| Variable | M(SD) | | t | p | Cohen's d |
| | Low Workload | High Workload | | | |
| Duty Cycle (longitudinal) | .47(.04) | .51(.06) | 5.51 | **< .001** | .92 |
| Duty Cycle (lateral) | .44(.04) | .45(.04) | 1.15 | .26 | .19 |
| PIW1 (longitudinal) | .14(.01) | .16(.02) | 4.99 | **< .001** | -.83 |
| PIW1 (lateral) | .14(.01) | .14(.02) | 1.76 | .09 | -.29 |
| Non-Parametric Data | | | | | |
| Variable | Mdn(IQR) | | Z | p | r |
| | Low Workload | High Workload | | | |
| Aggressiveness (longitudinal) | .00028(.00008) | .00029(.0001) | -3.40 | **< .001** | .57 |
| Aggressiveness (lateral) | .00041(.0001) | .00049(.0002) | -2.02 | .04 | .34 |

*Note.* Bolded numbers represent a statistically significant result with $p < .025$ (Bonferroni's adjustment of $\alpha = .05$). $M$ = mean, $Mdn$ = median, $SD$ = standard deviation, $IQR$ = interquartile range.

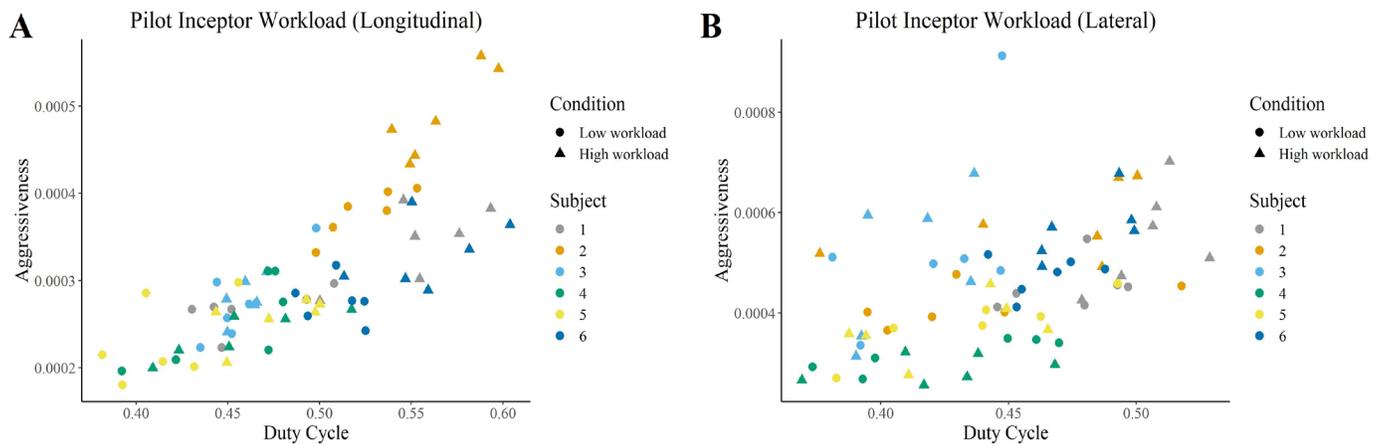

**Fig. 4.** Two-dimensional PIW plot of the (A) longitudinal stick data, and (B) lateral stick data.

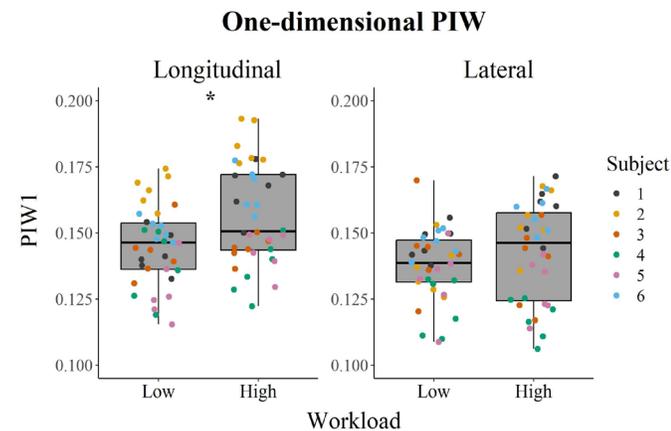

**Fig. 5.** Boxplots of longitudinal and lateral one-dimensional PIW (PIW1) per condition. Datapoints have been colored by participant. * $p < .001$.

*D. Predicting Workload with the Use of Pilot Inceptor Data*

The results of the binary logistic regression indicated that there was a positive association between longitudinal PIW1 and Workload, beta = 47.97, 95% CI [12.30, 88.87], $p < .05$. In other words, longitudinal PIW1 was predictive of the level of Workload in the current sample. There was no effect of lateral PIW1 on Workload, beta = -10.07, 95% CI [-49.34, 27.32], $p = .60$. The model's total explanatory power was weak, $R^2_{Tjur} = .11$.

IV. DISCUSSION

In the current study, we examined military pilots' stick behavior during a speed change maneuver in a VR flight simulator with and without an additional working memory task to vary the level of workload. We first compared Duty Cycle, Aggressiveness and one-dimensional Pilot Inceptor Workload (PIW) variables from longitudinal and lateral stick data in Low vs. High Workload conditions. Next, we trained a binary logistic regression model for prediction of workload from PIW values. We observed significantly higher PIW in the longitudinal stick data (i.e., pitch control) in the high workload trials when compared to the low workload trials. Furthermore, longitudinal PIW was predictive of workload, that is, a higher longitudinal PIW was associated with a larger probability of high workload. However, none of the lateral stick variables (i.e., roll control) differed between the two conditions and correspondingly lateral PIW was not predictive of workload.

Our findings regarding longitudinal stick data are in line with previous findings from low-fidelity simulator studies [12], [21], in which increased task difficulty (i.e., flying within the virtual boundaries of a given flight path of decreasing sizes, and adding secondary tasks to the initial flight maneuvers) increased Duty Cycle and Aggressiveness. As Gray [22] explains, when pilots are flying, they do not constantly change their control inputs. Instead, these inputs change as the flight's demands change. Despite this explanation, different findings were obtained for the lateral stick data.

Our different findings regarding the two directions of control input might be due to the specific task characteristics used in this study. A speed change requires mostly pitch input to compensate for the loss of lift and yaw input to counter the cross-coupling effect of propeller airflow on the vertical tail. This requires pilots to actively control the pitch attitude to maintain altitude, whereas yaw deviations can be compensated by using the pedals. Future work should evaluate whether our findings are observed in different flight tasks, such as turns, to determine whether the PIW obtained from the two directions of stick control input are task-dependent. It is expected that PIW will be affected in both directions for maneuvers that include both directions of control input, such as a climbing turn. Alternatively, as our task required closed-loop control, an open-loop approach in which pilots have to provide occasional control inputs and anticipate the response of the aircraft, [19], may also be explored. Overall, while the current study shows that PIW is predictive of workload in the speed change task, more experiments are necessary to determine whether this behavioral measure can predict workload in various flight tasks and training contexts.

It should be noted that the explanatory power of our model in the current study was weak. This may be due to the small sample size. Another limitation with respect to our model is that we used the entire duration of trials (210 sec per trial) to compute Duty Cycle and Aggressiveness, as opposed to extracting the part of trial that was strictly associated with the completion of the flight task. This could possibly have diminished the variability of PIW during the task, although participants were required to control and keep the virtual aircraft straight-and-level after completion of the primary flight task.

While behavioral measures reflect well on one's mental workload [35], it is expected that adding other predictors such as (neuro)physiological measures (e.g., EEG, pupil diameter change, etc.) would increase the explanatory power to predict the pilot's workload. Pupil diameter change has previously been associated with Duty Cycle and Aggressiveness derived from stick deflections in VR flight tasks [23]. Recently, EEG data have been implemented in an offline classification of workload of pilots during real flight [36]. Additionally, Kakkos et al. [37] have shown that EEG data can be used to classify workload of novice users in low-fidelity VR simulators using machine learning methods. However, no attempts have been made to classify workload of pilots using multi-modal data during VR flight training (in real-time), nor has this paradigm been used to create an adaptive VR flight training system for optimization of training [5].

Ultimately, we aim to predict a pilot's workload in real-time, and the current study indicated this possibility using PIW derived from controller's stick inputs. Using control input as a real-time indicator of workload is non-intrusive and does not require physiological sensors, which allows for the evaluation and personalization of pilot (VR) simulator training. Future studies should focus on how to increase the predictive power of the proposed models for prediction of pilots' workload, using advanced machine learning algorithms, additional measures of control input from various flight maneuvers as well as multimodal (neuro)physiological measurements.

The current study provides evidence that measuring PIW in a VR flight simulator yields a real-time and non-invasive means to determine workload and paves the way for more advanced (neurophysiological) techniques.